\documentclass[twocolumn,amsmath,showkeys,prl]{revtex4-1}

\usepackage{dcolumn}% Align table columns on decimal point
\usepackage{bm}% bold math
\usepackage[T1]{fontenc}
\usepackage{amsmath}
\usepackage{graphicx}

\usepackage{hyperref}
\hypersetup{colorlinks=true, linkcolor=blue, citecolor=red, urlcolor=magenta, pdftitle={bousquet}, pdfauthor={Eric Bousquet}}
\begin{document}
\title{Induced magnetoelectric response in \emph{Pnma} perovkites}

\author{Eric Bousquet$^{1,2}$}
\author{Nicola Spaldin$^{1}$}
\affiliation{$^1$ Department of Materials, ETH Zurich, Wolfgang-Pauli-Strasse 27, CH-8093 Zurich, Switzerland}
\affiliation{$^2$Physique Th\'eorique des Mat\'eriaux, Universit\'e de Li\`ege, B-4000 Sart Tilman, Belgium}

\begin{abstract}
We use symmetry analysis to show that the $G$, $C$ and $A$-type antiferromagnetic $Pnma$ perovskites can exhibit
magnetoelectric (ME) responses when a ferroelectric instability is induced with epitaxial strain.
Using first-principles calculations we compute the values of the allowed ME response in strained CaMnO$_3$ as a model system. 
Our results show that large linear and non-linear ME responses are present and can diverge when close to the ferroelectric phase transition.
By decomposing the electronic and ionic contributions, we explore the detailed mechanism of the ME response.
% we propose new routes to designing materials with large ME responses.
\end{abstract}

\maketitle

Interest in magnetoelectric (ME) materials has increased over the last few years because of their
cross coupling between the electric polarization and magnetization and their
consequent potential for technological applications \cite{fiebig2005}.
However, the search for good MEs is facing difficulties: compounds with the required symmetry (breaking 
of both the time and space inversion) are uncommon, and when these requirements are met, it is often 
at low temperatures. 
In addition, although the magnitude of the response is in principle bounded by
the product of the dielectric and magnetic permeabilities ($\alpha \le \sqrt{\epsilon \mu}$),
in practice it tends to be much smaller than this value.
One promising direction in the search for improved magnetoelectrics is the exploration of multiferroic 
materials, since the presence of multiple ferroic orders often presents the desired coupling properties 
for large ME responses \cite{fiebig2005, wojdel2009, wojdel2010}. 
Another route is the engineering of
artificial heterostructures with specific chemistries and symmetries \cite{bousquet2008,rondinelli2008,rondinelli2011}.
Here we demonstrate from symmetry considerations that the $Pnma$ $G$, $C$ or $A$-type antiferromagnetic 
perovskites, which are not multiferroic and do not allow a ME response in their bulk form, can become
ME when a polar distortion is induced using thin film heteroepitaxial strain.
Then, using first-principles calculations for a model example -- $Pnma$ CaMnO$_3$ --
we show that particularly large ME responses can be achieved in the vicinity of the ferroelectric phase transition.

\emph{Technical details}---
We performed all calculations within density functional theory as implemented in the VASP code \cite{vasp1,vasp2}.
Since the purpose of the present study is to provide a model example, we restricted ourselves to the Local Density Approximation (LDA) functional and intentionally avoided studying the $U$ and $J$ dependence of the LDA$+U$ functional \cite{bousquet2010b}.
We oriented the $Pnma$ unit cell with the longest axis along the $b$ direction, and
applied a cubic epitaxial strain on the $a$ and $c$ directions by imposing $a=c$ and relaxing the $b$ direction.
To compare with previous studies on strained CaMnO$_3$ \cite{bhattacharjee2009}, we performed all calculations at the 
calculated generalised gradient approximation (GGA) PBEsol volumes for each strain (see Supplemental Information).
The non-collinear properties and the ME response were well converged at a plane wave cutoff of 550 eV and a 4$\times$2$\times$4 k-point grid.
To compute the ME responses, we applied a Zeeman magnetic field to the spins with the spin-orbit coupling included and extracted the polarization response as described in Ref.~\onlinecite{bousquet2011}.

\emph{Group theory analysis}---
The $Pnma$ structure is obtained by condensation of three antiferrodistortive (AFD) instabilities coming from the $M$ and $R$ zone boundary points of the high symmetry $A$BX$_3$ cubic cell \cite{stokes2002}; 
The primitive unit cell contains four formula units.
When considering only one type of magnetic atom at the Wykoff site $4a$ in the $Pnma$ cell, four collinear magnetic orders are likely: ferromagnetic ($F$) and three antiferromagnetic (AFM) of $G$, $C$ and $A$ types.
For each collinear magnetic order we consider spin directions along $x$, $y$ or $z$, corresponding to the $a$, $b$ and $c$ axes of the crystal.
This allows us to define 3$\times$4 magnetic order parameters.
In Tab.~\ref{tab1} we report the symmetry characters of each of these magnetic order parameters according to their transformation under the symmetries of the $Pnma$ space group.
Interestingly, the magnetic order parameters group in threes with the same character.
In an expansion to second order of the energy with respect to these order parameters, the couplings between the three order parameters with the same character are allowed.
According to Tab.~\ref{tab1}, we then conclude that for $G$, $C$, $A$ or $F$ magnetic order in the $Pnma$ space group and for any easy axis, spin cantings in the two other directions are allowed.
We note that if we consider only the AFM magnetic orders, then three characters out of four allow $F$ spin canting \emph{i.e.} for weak ferromagnetism (FM).

\begin{table}[htbp!]
\begin{center}
\begin{tabular}{cccccc}
\hline
\hline
                      &   \multicolumn{2}{c}{\emph{Pnma}}   & & \multicolumn{2}{c}{\emph{Pmc2$_1$}} \\
                      &   Character & linear-ME  & & Character & linear-ME  \\
 \hline
$G_z$, $A_x$, $F_y$  & A$_g$	  &  $\oslash$ & & B$_1$	  & $\alpha_{xy, yx}$ \rule[-1ex]{0pt}{3.5ex} \\
$G_y$, $C_x$, $F_z$  & B$_{1g}$	  &  $\oslash$ & & A$_2$	  & $\alpha_{xx, yy, zz}$ \rule[-1ex]{0pt}{3.0ex} \\
$A_y$, $C_z$, $F_x$  & B$_{2g}$	  &  $\oslash$ & & B$_2$	  & $\alpha_{xy, yx}$ \rule[-1ex]{0pt}{3.0ex} \\
$G_x$, $C_y$,  $A_z$ & B$_{3g}$	  &  $\oslash$ & & A$_1$	  & $\oslash$ \rule[-1ex]{0pt}{3.0ex} \\
\hline
\hline
\end{tabular}
\caption{Symmetry character of the magnetic order parameters of the perovskite 20 atom unit cell with the \emph{Pnma} and \emph{Pmc2$_1$} space groups following the standard settings as given in the Bilbao Crystallographic Server.
When it is allowed, we report the components of the linear ME tensor that are non-zero.}
\label{tab1}
\end{center}
\end{table}

As shown in Tab.~\ref{tab1}, if we consider the condensation of any of the magnetic order parameters in the $Pnma$ structure, none of the resulting magnetic space groups permits for ME response.
This is because, although the time-reversal symmetry is broken, the inversion center is still preserved.
One possibility to break the inversion symmetry is with polar displacements.
Recently, it was shown from first-principles that it is possible to induce a ferroelectric (FE) instability in $Pnma$ CaMnO$_3$ with epitaxial strain \cite{bhattacharjee2009}.
When the polarization develops, the system condenses into the \emph{Pmc2$_1$} space group which has no inversion center.
In Tab.~\ref{tab1} we report the symmetry characters of the magnetic order parameters in the \emph{Pmc2$_1$} space group.
Interestingly, the magnetic orders stay in the same groups of three as in the $Pnma$ structure.
Three out of four characters in the new \emph{Pmc2$_1$} space group allow linear ME coupling. 
This means that in $G$, $C$ and $A$-type AFM $Pnma$ perovskites with one type of magnetic cation, it is possible to induce a ME response if a FE instability develops with epitaxial strain.

\emph{Ground state properties}---
Most previous studies on CaMnO$_3$ did not explore the possibility of non-collinear spin canting;
to our knowledge there is only one old experimental report of weak FM \cite{yudin1966}.
Therefore we first performed calculations for bulk CaMnO$_3$ including the spin-orbit coupling and explicitly checking all possible collinear and non-collinear magnetic ground states.
We found that the lowest energy magnetic ordering is $G$-type AFM with the easy axis along the $z$ direction. 
This indeed allows a canting of the spins of the $F$-type along the $y$ direction and of the $A$-type AFM along the $x$ direction ($G_zF_yA_x$ ground state), consistent with our group theory analysis reported in Tab.~\ref{tab1} and the experimental measurements \cite{yudin1966}.
The calculated canting angles are 2.6$^\circ$ along the $x$ direction and 1.0$^\circ$ along $z$, resulting in a weak FM of 0.04 $\mu_B$ per Mn atom.
This calculated weak FM is larger than the experimental report (0.004 $\mu_B$ \cite{yudin1966}); the discrepancy could be due to experimental uncertainty or our use of the LDA approximation.

\begin{center}
\begin{figure}[htbp!]
 \centering
 \includegraphics[width=7.7cm,keepaspectratio=true]{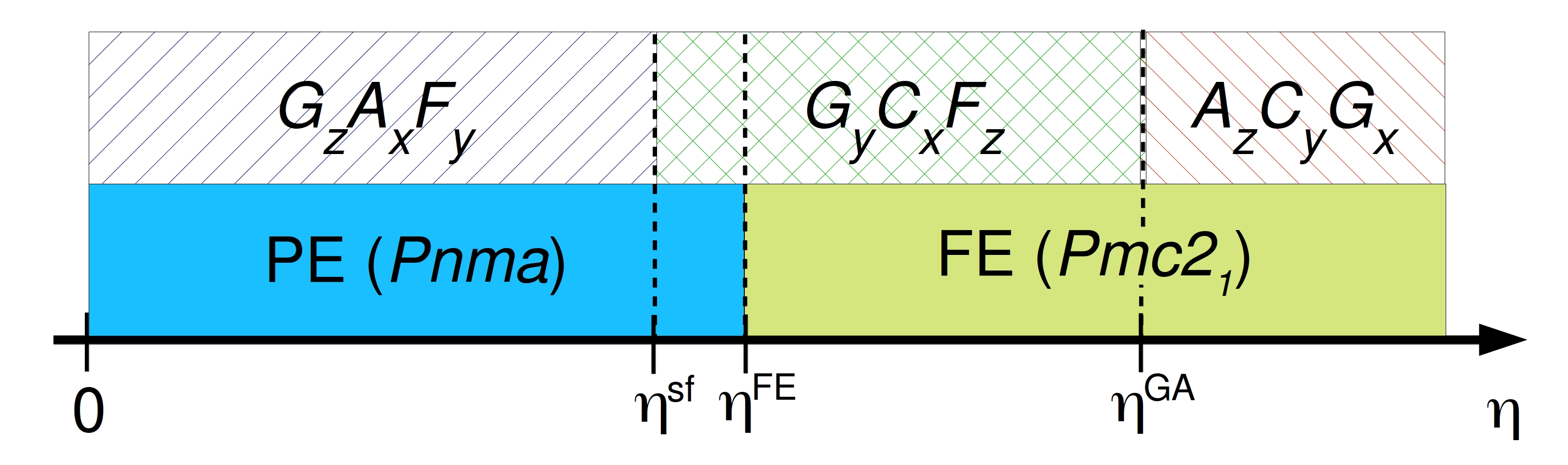}
 \caption{Multiferroic phase diagram of CaMnO$_3$ under tensile epitaxial strain.
The top part shows the magnetic orders and the lower part shows the crystallographic orders (PE = paraelectric).
$\eta^{sf}=$ 2.6\%, $\eta^{FE}=$ 3.2\% and $\eta^{GA}=$ 4.6\%.
Between $\eta=0$ and $\eta^{sf}$ the magnetic point group is $m'mm'$, between $\eta^{sf}$ and $\eta^{FE}$ it is $m'm'm$, between $\eta^{FE}$ and $\eta^{GA}$ it is $m'm'2$ and beyond $\eta^{GA}$ it is $mm2$.}
 \label{fig1:phasediag}
\end{figure}
\end{center}

Having verified the accuracy of our calculated ground state properties, we then checked how these properties are affected by the epitaxial strain and by the development of the FE polarization.
To determine the lowest energy phase, we performed full atomic and spin relaxations at different epitaxial strains and for all of the magnetic order parameters reported in Tab.~\ref{tab1}.
We summarize the results in the phase diagram of Fig.~\ref{fig1:phasediag}.
As reported previously \cite{bhattacharjee2009}, we oberve that a FE instability appears beyond critical tensile epitaxial strain $\eta^{FE}$.
$\eta^{FE}$ was predicted previously within the GGA Wu-Cohen approximation  to be 2.0\%, the LDA
% **IS THAT PREVIOUS STATEMENT CORRECT?**
value obtained here is larger (3.2\%). 
This shows that the magnitude of strain required to induce ferroelectricty is strongly dependent on the approximations 
used (exchange-correlation functional, plane wave scheme, \emph{etc.}) in a calculation, and exact quantitative predictions 
should be made with caution. 
The electric polarization develops along the in-plane $c$ direction and increases with $\eta$, reaching large values at large strains (27 $\mu$C.cm$^{-2}$ at $\eta$=4.5\%).
Interestingly, we found that the amplitudes of the AFD rotations are only slightly modified (few \%) by the epitaxial strain and persist even when the polarization develops.

Looking at the magnetic properties we found that the strain induces a spin flop of the easy axis from the $z$ direction to the $y$ direction at a critical strain $\eta^{sf}$ = 2.6\% (Fig.~\ref{fig1:phasediag}) smaller than $\eta^{FE}$.
This transition keeps the primary $G$-type AFM magnetic order but, since the easy axis is modified, changes the type of spin canting allowed from $G_zF_yA_x$ to $G_yC_xF_z$.
With $G_y$ order, canting of the $C$-type along the $x$ direction and a weak FM along the $z$ directions are allowed (Tab.~\ref{tab1}).
When the FE transition takes place, we find no change of the magnetic ground state, with the system staying in the $G_yC_xF_z$ magnetic state.
We observe however that as the polarization increases, the total magnetic moment and spin cantings decrease by respectively a
few \% and a factor of three (between $\eta$ = 3.2\% and 4.5\%).
We confirm that this is an effect of the FE polarization and not of the strain by constraining the polarization to be 
zero by symmetry -- in this case we do not observe such a reduction of the spin cantings.
At large tensile epitaxial strain, we observe a magnetic phase transition from the $G$-type AFM to the $A$-type AFM with again a change of the easy axis from the $y$ ($G_y$) to the $z$ direction ($A_z$) at $\eta^{GA}$ = 4.6\%.
Here again, consistent with Tab.~\ref{tab1}, with the $A_z$ order we observe spin cantings of the $C$-type along the $y$ direction and of the $G$-type along the $x$ direction ($A_zC_yG_x$).
The weak FM is then lost during this phase transition.

\begin{figure}[htbp!]
\begin{center}
 \includegraphics[width=7.2cm,keepaspectratio=true]{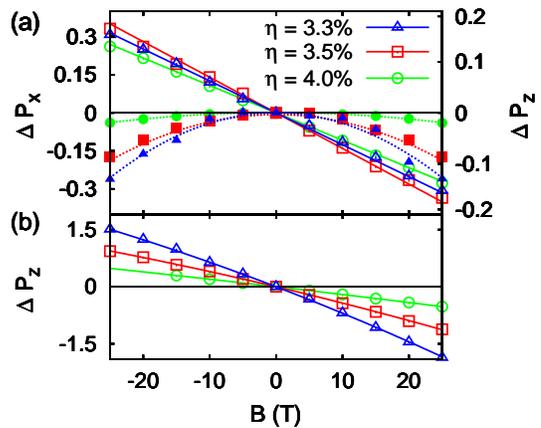}
\caption{Polarization response of CaMnO$_3$ under magnetic field at three different epitaxial strains. 
(a) Variation of polarization along the $x$ direction (empty symbols, plain lines) and along the $z$ direction (plain symbols, dashed lines) when the field is applied along $x$ and (b) Variation of polarization along the $z$ direction when the field is applied along $z$.}
\label{fig2:PvsB}
\end{center}
\end{figure}

\emph{ME response}---
From the phase diagram of Fig.~\ref{fig1:phasediag} and according to the group theory analysis of Tab.~\ref{tab1} we can see that in the region between $\eta^{FE}$ and $\eta^{GA}$ the system allows for a diagonal linear ME response ($G_yC_xF_z$ magnetic state in the \emph{Pmc2$_1$} crystal space group).
However, such analysis does not allow us to predict the amplitude of the response.
To determine the amplitude of the ME response, we performed calculations under a Zeeman magnetic field along the $x$, $y$ and $z$ directions and calculated the induced electric polarization \cite{bousquet2011}. 
In Fig.~\ref{fig2:PvsB} we report the induced change in polarization versus magnetic field at three different epitaxial strains between $\eta^{FE}$ and $\eta^{GA}$.
Fig.~\ref{fig2:PvsB}.a shows the variation of the polarization ($\Delta P$) along the $x$ direction ($\Delta P_x=P_x$ since $P_x=0$ at $B=0$) when a magnetic field is applied in the same direction, giving the $\alpha_{xx}$ component of the ME response.
We find a linear evolution of the polarization for $B_x$ between -25 and 25 T and ME response values: $\alpha_{xx}=-16$ ps.m$^{-1}$ at $\eta=3.3$\% and $-12$ ps.m$^{-1}$ at $\eta=4.5$\%.
These values are large compared with the prototypical ME compound Cr$_2$O$_3$ where the calculated ME response is 1.45 ps.m$^{-1}$ \cite{bousquet2011}.
In addition to the polarization induced along the $x$ direction, we also observe a change in the FE polarization along the $z$ direction even if the field is applied along the $x$ direction.
We report this response in Fig.~\ref{fig2:PvsB}.a where we see that it is highly non-linear.
This is in agreement with group theory wich shows that linear response $\alpha_{xz}$ is not permitted, but that the next order non-linear response $\beta_{xxz}$ 
\footnote{In our notation for $\beta_{ijk}$, the index $i$ and $j$ refer to the direction of the magnetic field and the index $k$ is for the direction of the polarization. We use $SI$ units \cite{rivera1994}} is allowed.
This non-linear response is small with respect to the linear response and is extremely sensitive to the epitaxial strain ($\beta_{xxz}=$ -7$\times$10$^{-19}$ s/A at 3.3\%).
In Fig.~\ref{fig2:PvsB}.b we report the variation of polarization along the $z$ direction when a magnetic field is applied in the same direction ($\alpha_{zz}$ component).
Here again the polarization response is strongly sensitive to the epitaxial strain and deviates from an ideal linear response, consistent with symmetry analysis which yields a non-zero $\alpha_{zz}$ and $\beta_{zzz}$.
The $\alpha_{zz}$ value is particularly large near to the FE transition ($\alpha_{zz}=-85$ ps.m$^{-1}$ at $\eta=$ 3.3\%, close to $\eta^{FE}$) and decays rapidly away from $\eta^{FE}$ ($\alpha_{zz}=-19$ ps.m$^{-1}$ at $\eta=$ 4.5\%) and $\beta_{zzz}$ is sizeable ($-4.2\times10^{-16}$ s/A at 3.3\%).
Finally, when applying a magnetic field along the $y$ direction, we do not observe a polarization along the $y$ direction.
This shows that even when a component is symmetry allowed, it can be extremely small in amplitude, here lower than the precision of our calculations.
With the field applied along $y$, we observe however a tiny polarization response along the $z$ direction which corresponds to the allowed non-linear ME tensor component $\beta_{yyz}$.

\begin{figure}[htbp!]
\begin{center}
\includegraphics[width=9.7cm,keepaspectratio=true]{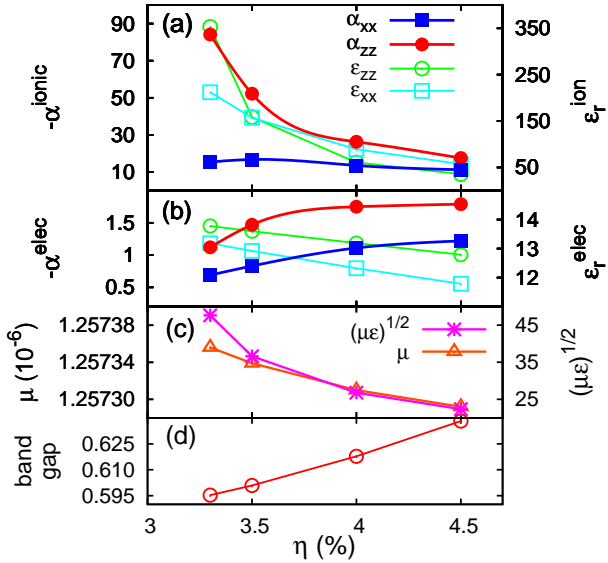}
\caption{$xx$ (squares) and $zz$ (circles) components of the ME (plain symbols, in ps/m) and dielectric constant (empty symbols, SI units) versus epitaxial strain: 
(a) ionic contribution, (b) electronic contribution. (c) magnetic permeability (triangles, SI units) and ME upper bound $\sqrt{\varepsilon\mu}$ (stars, in ns/m whith $\varepsilon=\varepsilon_0(\varepsilon_r^{ion}+\varepsilon_r^{elec})$) versus epitaxial strain. (d) electronic bang gap (eV).
}
\label{fig3:MEvsStrain}
\end{center}
\end{figure}

\emph{Electronic versus ionic contribution}---
Recently it has been shown by explicit calculation of the Zeeman electronic contribution to the ME response ($\alpha^{elec}$) in Cr$_2$O$_3$ and LiNiPO$_4$, that $\alpha^{elec}$ can be comparable in magnitude to the ionic contribution ($\alpha^{ionic}$) \cite{bousquet2011}.
It has also been suggested that in FE perovskites such as BiFeO$_3$, the ME response is dominated by the FE soft mode (ionic contribution) \cite{iniguez2008,wojdel2009,wojdel2010}.
To check how large are the electronic and ionic contributions to the ME response in strained CaMnO$_3$, we computed these two contributions as in Ref.\cite{bousquet2011}.
We summarize the results in Fig.~\ref{fig3:MEvsStrain}.a and b, where the ionic and electronic contributions to $\alpha_{xx}$ and $\alpha_{zz}$ are plotted with respect to the epitaxial strain.
We can conclude the following:
First, the ionic contribution clearly dominates the total response.
$\alpha^{ionic}_{zz}$ is extremely sensitive to the strain and diverges when approaching the FE phase transition, consistent with the softening of a polar mode along the $z$ direction \cite{wojdel2010}.
$\alpha^{ionic}_{xx}$ is however much less sensitive to the strain and does not show any divergence when approaching the FE transition.
We find that the response of $\alpha_{xx}$ is mainly dominated by a relatively soft mode (110 cm$^{-1}$) which keeps almost the same frequency for epitaxial strain  from $\eta^{FE}$ to $\eta^{GA}$.
Second, the electronic contribution (Fig.~\ref{fig3:MEvsStrain}.b) shows the opposite trend to the ionic contribution in both $\alpha_{xx}$ and $\alpha_{zz}$.
While the ionic contribution has the tendency to decrease when the strain increases, the electronic contribution increases.
We remark that, even though the electronic contribution is much smaller than the ionic contribution, its absolute value is large compared with the values reported for Cr$_2$O$_3$ ($\alpha^{elec}=$ 0.34 ps.m$^{-1}$) and LiNiPO$_4$ ($\alpha^{elec}=$ 1.1 ps.m$^{-1}$) \cite{bousquet2011}.
In strained CaMnO$_3$ the electronic contribution alone can even be larger than the total ME response of Cr$_2$O$_3$ ($\alpha^{tot}=$ 1.45 ps.m$^{-1}$ \cite{bousquet2011}).

The opposite trend between $\alpha^{ionic}$ and $\alpha^{elec}$ shows clearly that the electronic response is not driven by a phonon instability.
We can make an analogy with the dielectric permittivity in crystals ($\varepsilon_r$).
The ionic contribution to $\varepsilon$ is directly affected by the softening of polar phonon modes and is responsible for the divergence of $\varepsilon_r$ close to a FE phase transition.
However the electronic contribution (or high frequency $\varepsilon_r^\infty$) does not diverge at a FE phase transition, but rather diverges close to an electronic instability such as a metal-insulator (M-I) phase transition.
We expect similar behaviour for $\alpha^{elec}$, however we are unable to test it in strained CaMnO$_3$ as it does not show a M-I transition.
We can, nevertheless, comment on the dependence of $\alpha^{elec}$ on the gap.
The magnetic phase transition between $G_yC_xF_z$ and A$_z$C$_y$G$_x$ at $\eta^{GA}=$ 4.6\% is first order and $\mu$, $\alpha^{elec}$'s and $\varepsilon^{elec}_r$'s reported in Fig.~\ref{fig3:MEvsStrain}.b do not show diverging behaviour.
While the decrease of $\varepsilon_r^\infty$ is directly related to the increase of the band gap (see Fig.\ref{fig3:MEvsStrain}.d), $\alpha^{elec}$ in fact \emph{increases} with decreasing gap (See Fig.~\ref{fig3:MEvsStrain}.b).
The magnetic permeability $\mu$ (we have here $\mu_{xx}=\mu_{zz}=\mu^{elec}=\mu$) decreases in the same range of strain (Fig\ref{fig3:MEvsStrain}.c). 
This shows that the link between $\alpha^{elec}$ and $\varepsilon_r^\infty$, $\mu$  and the band gap is not straightforward. 
While a formulation of the orbital contribution to $\alpha$ \cite{malashevisch2011} and a generalized Lyddane-Sachs-Teller relationship between $\alpha$, $\mu$ and $\varepsilon$ in ME \cite{resta2011} have been reported, an exact formulation of Zeeman $\alpha^{elec}$ is still needed.
In Fig.\ref{fig3:MEvsStrain}.c we also report the upper bound of $\alpha$ which is equal to $\sqrt{\varepsilon\mu}$ \cite{brown1968}.
As we can see $\sqrt{\varepsilon\mu}$ can be as large as 50 ns/m while $\alpha$ is of the order of 90 ps/m.
$\alpha$ is then about 0.1 \% of its upper bound as reported for Cr$_2$O$_3$ \cite{brown1968}.

From these observations we propose the following design rules to obtain large ME responses: (i) one or more soft polar modes to increase $\alpha$ through $\varepsilon_r^{ion}$, (ii) a magnetic instability to increase $\alpha$ through $\mu$ (iii) proximity to an electronic instability such as a M-I phase transition to increase $\varepsilon_r^\infty$.
(i) is clearly evidenced in our results on strained CaMnO$_3$.
(ii) is not observed in spite of the magnetic phase transition at $\eta^{GA}$.
This is because the magnetic phase transition at $\eta^{GA}$ is of the first order which does not result to the divergence of $\mu$.
However, second order magnetic phase transitions should have the diverging effect on $\mu$ and $\alpha$.
(iii) will cause $\varepsilon_r^\infty$ to diverge but we cannot yet conclude how it will affect $\alpha$ and further explorations are needed to clarify this point.
If (i), (ii) and (iii) can be achieved simultaneoulsy, a very large ME response could be obtained if the effects on $\alpha$ are of the same sign.
While in this paper we have shown the divergence of the ME response with epitaxial strain, the effect will also occur at phase transitions induced by temperature, pressure, \emph{etc}.

We emphasise that our findings are not restricted to the case of CaMnO$_3$, but are valid for all $G$, $C$ and $A$-type 
AFM $Pnma$ perovskites since the group theory analysis reported in Tab.~\ref{tab1} is valid for any $A$, $B$ and $X$ ions.
Since the $Pnma$ structure is the most common natural ground state of the $A$BX$_3$ perovskites \cite{howard2004}, this offers many possibilities for creating new ME materials with epitaxial strain.
Furthermore, mixing chemistries in superlattices vastly increases the possibility of generating phases with coexisting
phonon, electronic and magnetic instabilities and giant ME responses \cite{bousquet2008, rondinelli2008, rondinelli2011}.

We thanks K. Delaney, V. Gopalan and A. Scaramucci for fruitful discussions.
This work was supported by the ETH Z\"{u}rich and FRS-FNRS Belgium (EB).

%   \bibliography{biblio}
%Merlin.mbs v4.21 2009-07-09.
%Merlin.mbs v4.21 2009-07-09.
%

\end{document}